\documentclass{article}
\usepackage{amssymb}
\usepackage{epsfig}
\usepackage{psfig}
\usepackage{latexsym}

\begin{document}

\begin{flushright}
   MZ-TH/02-31  \\
   FTUV-0412/02 \\
   December 2002 \\
\end{flushright}

\medskip

\begin{center}
{\LARGE Bottom Quark Mass and QCD Duality\footnote{{\LARGE {\footnotesize %
Supported by MCYT-FEDER under contract FPA2002-00612, EC-RTN under contract
HPRN-CT2000/130 and partnership Mainz-Valencia Universities.}}}}\vspace{2cm}

{\large J. Bordes}$^{a}${\large , J. Pe\~{n}arrocha}$^{a}${\large \ and K.
Schilcher}$^{b}\bigskip $

$^{a}${\normalsize Departamento de F\'{\i}sica Te\'{o}rica-IFIC, Universitat
de Valencia}

{\normalsize E-46100 Burjassot-Valencia, Spain}

$^{b}${\normalsize Institut f\"{u}r Physik, Johannes-Gutenberg-Universit\"{a}%
t}

{\normalsize D-55099 Mainz, Germany\vspace{2cm}}

\textbf{Abstract\medskip }
\end{center}

\begin{quote}
The mass of the bottom quark is analyzed in the context of QCD finite energy
sum rules. In contrast to the conventional approach, we use a large momentum
expansion of the QCD correlator including terms to order $\alpha
_{s}^{2}(m_{b}^{2}/q^{2})^{6}$ with the upsilon resonances from $e^{+}e^{-}$
annihilation data as main input. A \ stable result $m_{b}(m_{b})=(4.19\pm
0.05)\,\,\mathrm{GeV}$ for the bottom quark mass is obtained. This result
agrees with the independent calculations based on the inverse moment
analysis.

\bigskip
\end{quote}
PACS: 12.38.Bx, 12.38.Lg.
{\newpage }

\vspace*{2cm}

\section{Introduction}

With the Standard Model being so well established the major theoretical
effort of today's theoretical and experimental investigations goes into a
precise determination of the parameters of the model. An important example
is the determination of the CKM matrix elements and their relative phases in
current B-physics experiments. A better knowledge of these parameters will
lead to a better understanding of the structure of weak interaction
including CP violation, and possibly the discovery of "new physics". For an
analysis of most of the experiments in the B-sector a precise value of the
bottom quark mass is essential.

In the last few years, there has been increased activity with the aim of
determining the bottom quark mass $[1-10]$ to higher accuracy. 
The most popular method was the one based global duality in QCD. In
so-called inverse moment sum rules derivatives of the two point correlation
function of the b quark vector current are compared with experimental
information from electron-positron annihilation into hadrons \cite{NdeR}.
These sum rules seem to be a suitable method since it is very sensitive to 
the heavy (b-quark) quark mass.
Recent analysis using inverse moments involving 10 to 20 inverse powers of
the squared energy can be found in \cite{NarisPL,PICH} and, using a lower
power inverse moments, in \cite{KUHN}. Non relativistic QCD, potential
models (for a review see \cite{HOANG}) and lattice techniques \cite%
{lattice} have also been considered in order to determine the heavy quark
masses.
However, in the case of the lattice techniques and potential models, there
exist problems with the choice of the mass definition (running-,
pole-, threshold mass) and the size of non-perturbative effects. So far, the
available experimental information relies mainly on the knowledge of the
masses and widths of six upsilon resonances, three of them lying below the
continuum threshold of the $B\overline{B}$ meson production.

The method we propose here is somewhat orthogonal and complementary to the
one based on the inverse moment sum rules: we employ on the theoretical side
the large momentum expansion of QCD, i.e. an expansion in m$_{b}^{2}/s$,
where s is the square of the CM energy. Such an expansion makes sense as
long as $s$ is far enough away from the continuum threshold. We will
consider the perturbative expansion up to second order in the strong
coupling constant and twelve powers of the b-mass over energy ratio using
the results of the reference \cite{Chet1}. An expansion to the third order in
the strong coupling but with only four powers in the mass-energy ratio is
also known \cite{Chet2}, which we will not make use of for reasons of
consistency. On the phenomenological side of the sum rule we will consider
the six upsilon resonances. In addition, our method should be sensitive to
the poorly known continuum data. We circumvent this problem by a judicious
use of quark-hadron duality. Above the resonance region, where experimental
data are poorly known, we incorporate a real polynomial in the sum rule \cite%
{ACD} or, what is the same, a suitable combination of positive moments, in
such a way that the contribution of the data in the region above the
resonances can be practically eliminated. The method has been successfully
checked in the charm sector \cite{PS}, where the charm quark mass was
predicted by using the experimental data and the polynomial insertion in the
intermediate region. The result for the charm mass was found in good
agreement with the ones obtained using other methods. Of course, in the
b-sector one is in a truly heavy quark regime and it is not clear that the
same approach may lead to a result for the b-mass of competitive accuracy.
We will show that this is actually the case.

The plan of this note is the following: in the second section we briefly
review the theoretical method used here which is based on weighted finite
energy sum rules, in the third section we discuss the theoretical and
experimental inputs used in the calculation and present our results for the
bottom quark mass with a discussion of the errors. We finish the paper giving
the conclusions.

\section{The calculation}

In order to write down the sum rule relevant to our case, we apply Cauchy's
theorem to the two point correlation function $\Pi (s)$, with the
appropriate flavor content in the quark currents, and weight the integration
with a polynomial $P(s)$. The integration path extends along a circle of
radius $\left\vert s\right\vert =s_{R}$, and along both sides of the
physical cut $\left[ s_{0},s_{R}\right]$. The polynomial $P(s)$ does not
change the analytical properties of $\Pi (s)$ in the integration region, so
we obtain the following sum rule 
\begin{equation}
\frac{1}{\pi }\int_{s_{0}}^{s_{R}}P(s)\,\mathrm{Im}\Pi (s)\,\,ds=-\frac{1}{%
2\pi i}\oint_{\left\vert s\right\vert =s_{R}}P(s)\,\,\Pi (s)\,\,ds,
\label{SR}
\end{equation}%
where $s_{0}$ is the physical threshold of the $b\overline{b}$ channel,
starting at the first $b\overline{b}$ resonance below the continuum
threshold $s_{C}\geqslant s_{0}$. The integration radius $s_{R}$ is chosen
in such a way that on the circle the asymptotic expansion of QCD, $\Pi _{%
\mathrm{QCD}}(s)$, constitutes a good approximation to the two point
correlator. The left hand side of equation (\ref{SR}), is related to the
experimental $e^{+}e^{-}$ annihilation cross section via the unitarity
relation, 
\begin{equation}
R(s;\,e^{+}e^{-}\rightarrow \mathrm{hadrons})=12\pi \sum_{\mathrm{flavors}%
}Q_{f}^{2}\,\,\mathrm{Im}\Pi (s).  \label{RATIO}
\end{equation}

\bigskip

In order to perform analytically the contour integration in (\ref{SR}) we take,
as a matter of convenience, the two point correlation function $\Pi
_{\mathrm{QCD}}(q^{2})$ which has been
calculated in \cite{Chet1} as an expansion up to second order (three
loops) in the strong coupling constant $\alpha _{s}$ and as a power series
of $m^{2}/q^{2}$ up to the sixth order, 
\begin{equation}
\Pi _{\mathrm{QCD}}(q^{2})=\Pi ^{(0)}(q^{2})+\left( \frac{\alpha _{s}(\mu )}{%
\pi }\right) \Pi ^{(1)}(q^{2})+\left( \frac{\alpha _{s}(\mu )}{\pi }\right)
^{2}\Pi ^{(2)}(q^{2}),  \label{QCD}
\end{equation}%
where the different terms of the expansion in $\alpha _{s}$ have the
form: 
\begin{equation}
\Pi ^{(i)}(q^{2})=\sum_{j=0}^{6}\sum_{k=0}^{3}A_{j,k}^{(i)}(m,\mu )\left(
\ln \frac{-q^{2}}{\mu ^{2}}\right) ^{k}\left( \frac{m^{2}}{q^{2}}\right)
^{j}.  \label{EXPAN}
\end{equation}%
In equations (\ref{QCD},\ref{EXPAN}) $\mu $ is the renormalization scale of
the perturbative calculation, $m$ is the running quark mass and $%
A_{j,k}^{(i)}(m,\mu )$ are the coefficients of the QCD perturbative
expansion which depend on $m$ and $\mu $ through powers of $\ln m^{2}/\mu
^{2}$ \cite{Chet1}. In the asymptotic expansion of equation (\ref{EXPAN})
one should include the non-perturbative terms due to the quark and gluon
condensates, which are known up to order $\alpha _{s}$ \cite{Gcond}. Their
contribution, however, is completely negligible in the range of energies of
interest to us. For this reason we will discard them from now on.

\bigskip

As commented in the introduction, the experimental side of the sum rule is
dominated by the six well established $b\overline{b}$ upsilon resonances.
With this experimental information in the narrow width approximation we
have the hadron cross section of equation (\ref{RATIO}) given by 
\[
R(s)=\frac{9\pi }{\alpha _{em}^{2}}\sum_{\mathrm{res}}M_{\mathrm{res}%
}\,\,\Gamma _{\mathrm{res}}\,\,\delta (s-M_{\mathrm{res}}^{2}),%
\]
where $M_{\mathrm{res}}$ and $\Gamma _{\mathrm{res}}$ are respectively the
masses and electronic widths of the six resonances, and $\alpha
_{em}=1/131.8 $ is the electromagnetic coupling constant taken at the
typical scale of $10$ GeV, where the resonances are produced.

\bigskip

Furthermore, in the theoretical side of the sum rule, equation (\ref{SR}), $%
P(s)$ is taken to be a third degree polynomial 
\[
P(s)=a_{0}+a_{1}s+a_{2}s^{2}+a_{3}s^{3},
\label{POLY}
\]%
whose coefficients are fixed by imposing a normalization condition $%
P(s_{0})=1$, and requiring that it should minimize the contribution of the
continuum $\left[ s_{C},s_{R}\right] $ in a least square sense, i.e., 
\[
\int_{s_{C}}^{s_{R}}s^{n}P(s)\,\,ds=0\,\,\mathrm{for}\,\,n=0,1,2,%
\label{CONDIT}
\]%
where $s_{C}$ is the value of the continuum physical threshold for the $B%
\overline{B}$ meson production. The choice of this polynomial guarantees
that the contribution of the experimental data in the continuum region
(which is experimentally poorly known) will be negligible in the sum rule as
compared to the contribution of the lower resonances below the continuum
threshold. In fact, a smooth continuum contribution which can be described
by a second order polynomial vanishes identically. In this way, we avoid the
inaccurate experimental information in the $b\overline{b}$ continuum region.
This approximation procedure was checked in the $c\overline{c}$ channel,
where there is good experimental information on the continuum due to recent
results from the BES II collaboration \cite{bes}. Accurate and consistent
results for the charm mass have been obtained, by either incorporating the
continuum or eliminating it by the procedure above \cite{PS}.

\bigskip

Notice also that the dependence on the method, which relies on the choice of
the
polynomial (\ref{POLY}), is very weak. To check that fact we have performed the 
same calculation using polynomials of different degrees as well as changing the
boundary conditions of the third degree polynomial (\ref{CONDIT}), although 
keeping the same philosophy as commented in the paragraph above.
For instance for a two degree polynomial we
find a difference of .01 Gev in the bottom mass but the experimental error
coming from the uncertainties in the resonance data is double than the one we
find with a degree three polynomial. On the other hand a polynomial of three
degree changing slightly the boundary conditions gives just the same
result. 

Higher order polynomials, as well as the inclusion of the incomplete order 
$\alpha^3_s$ in equation (\ref{QCD}) can be considered as well. However, before 
doing so, a detailed study of the different moments involved in equation 
(\ref{SR}) should be performed. We defer this study to a further publication.

With our study we conclude that our selection for the polynomial do not
preclude the results found nor the errors presented in this work.

\bigskip

The integrals that we have to calculate on the right hand side of the sum
rule, equation (\ref{SR}), are 
\[
J(p,k)=\frac{1}{2\pi i}\oint_{\left\vert s\right\vert =s_{R}}s^{p}\left( \ln 
\frac{-s}{\mu ^{2}}\right) ^{k}ds,
\]%
for $k=0,1,2,3$ and $p=-6,-5,..,3$. Their analytic evaluation can be found
in reference \cite{PPSS}. After integration, equation (\ref{SR}) can be cast
into the form 
\begin{eqnarray}
&&9\pi (131.8)^{2}\sum_{res}M_{res}\,\,\Gamma _{res}\,\,P(M_{res}^{2}) 
\nonumber \\
&=&-\frac{4\pi ^{2}}{3}\sum_{n=0}^{3}\sum_{i=0}^{2}\sum_{j=0}^{6}%
\sum_{k=0}^{3}a_{n}\left( \frac{\alpha _{s}}{\pi }\right)
^{i}A_{j,k}^{(i)}(m,\mu )\left( m^{2}\right) ^{j}J(n-j,k),  \label{KEYEQ}
\end{eqnarray}%
giving a non-linear equation with the quark mass ($m$) as the only unknown.
To proceed further and solve this equation we have to choose a suitable
value of $s_{R}$ for which perturbative QCD gives a good approximation for
the correlation function. Furthermore, the scale $\mu $ can be chosen
arbitrarily to give a prediction of the quark running mass at this scale.
For calculational convenience, we take $\mu ^{2}=s_{R}$ since most of the
logarithmic terms vanish after integration. Then, after solving equation (%
\ref{KEYEQ}), we run the quark mass from that scale to the mass scale
itself, $m(\mu =m)$, by means of the corresponding renormalization group
equations at the appropriate loop order in the strong coupling constant \cite%
{Santa,Larin97}.

\bigskip

As far as the theoretical input matches the experimental data within the
error bars, the results should be independent of the choice for $s_{R}$.
Nevertheless, the lack of experimental data in the continuum region of the $b%
\overline{b}$ channel makes sensible in our method to study the $s_{R}$
dependence of the quark mass results. Although this dependence is going to
be very small, due to the choice of our polynomial in the sum rule, we will
take
the quark mass prediction at the most stable value with $s_{R}$.

\section{Results}

The experimental inputs in our calculation are as follows:

\bigskip Firstly, the physical threshold $s_{0}$ is the squared mass of the
first low lying resonances in the $b\overline{b}$ channel, $\Upsilon (9460)$%
, 
\[
s_{0}=m_{\Upsilon }^{2}=9.46^{2}\,\,\mathrm{GeV}^{2},
\]%
whereas the continuum threshold $s_{C}$ is taken at the squared energy of
the $B\overline{B}$ meson production 
\[
s_{C}=(2m_{B})^{2}=(10.58)^{2}\,\,\mathrm{GeV}^{2}.
\]

Secondly, the absorptive part of the two point correlation function is
obtained from the six known $b\overline{b}$ vector resonances $\Upsilon(1S)$%
,..,$\Upsilon(6S)$ with the following masses and electromagnetic widths \cite%
{pdg,alb}.
\begin{equation}
\begin{tabular}{l}
$M(1S)=(9460.30\pm 0.26) \,\, \mathrm{MeV}$, \\ 
$M(2S)=(10023.26\pm 0.31)\,\, \mathrm{MeV}$, \\ 
$M(3S)=(10355.2\pm 0.5)\,\, \mathrm{MeV}$, \\ 
$M(4S)=(10580.0\pm 3.5)\,\, \mathrm{MeV}$, \\ 
$M(5S)=(10865\pm 8)\,\, \mathrm{MeV}$, \\ 
$M(6S)=(11019\pm 8)\,\, \mathrm{MeV}$,%
\end{tabular}
\begin{tabular}{l}
$\Gamma(1S)=(1.32\pm 0.05) \,\, \mathrm{keV}$, \\ 
$\Gamma(2S)=(0.520\pm 0.032)\,\, \mathrm{keV}$, \\ 
$\Gamma(3S)=(0.48\pm 0.04)\,\, \mathrm{keV}$, \\ 
$\Gamma(4S)=(0.248\pm 0.031)\,\, \mathrm{keV}$, \\ 
$\Gamma(5S)=(0.31\pm 0.07)\,\, \mathrm{keV}$, \\ 
$\Gamma(6S)=(0.130\pm 0.030)\,\, \mathrm{keV}$.%
\end{tabular}
\label{RDATA}
\end{equation}

Finally, for the strong coupling constant we take the input value at the
mass of the electroweak $Z$\ boson \cite{pdg} 
\begin{equation}
\alpha _{s}(M_{Z})=0.118\pm 0.003  \label{ALPHA}
\end{equation}%
and run down to the scale of the contour radius $s_{R}$ using the four loop
formulae of reference \cite{Santa}.

Now we proceed as follows. For $\mu ^{2}=s_{R}$ with $s_{R}$ in the range $%
\left[ 150,350\right] \,\,\mathrm{GeV}^{2}$ we determine $m_{b}^{(i)}(\mu )$
by solving equation (\ref{KEYEQ}) keeping terms in the perturbative
expansion up to order $(\alpha _{s}/\pi )^{i}$ for $i=0,1,2$. Then we run
the results from $\mu ^{2}=s_{R}$ to $\mu =m_{b}$ with the appropriate
renormalization group equations \cite{Larin97}. Finally, we choose the value of
$%
s_{R}$ which is most stable in the range of energies considered. In this way
we find stability at $s_{R}=240\,\,\mathrm{GeV}^{2}$ and the following
results for the bottom mass: 
\begin{eqnarray}
m_{b}^{(0)}(m_{b}) &=&4.09\,\,\mathrm{GeV},  \nonumber \\
m_{b}^{(1)}(m_{b}) &=&4.23\,\,\mathrm{GeV},  \label{mass} \\
m_{b}^{(2)}(m_{b}) &=&4.19\,\,\mathrm{GeV}.  \nonumber
\end{eqnarray}

To estimate the various errors arising in the calculation of the $b$ quark
mass we consider first the uncertainties in the masses and widths of the
resonances in equation (\ref{RDATA}). This gives an experimental error for
the mass $\varepsilon _{\mathrm{EXP}}=0.02\,\,\mathrm{GeV}$\footnote{Recently,
it has been claimed an enhancement of the experimental data in the $4S$ and
$5S$ resonance region \cite{pdgflaw}. This could change the final value of the
bottom mass by less than $0.01$ GeV, which is within the exprimental errors of
the method.}. Secondly, we
consider the uncertainty in the strong coupling constant, equation (\ref%
{ALPHA}), which leads to $\varepsilon _{\alpha }=0.01\,\,\mathrm{GeV}$.
Finally, we include a conservative asymptotic error originating from the
higher orders in expansion of equation (\ref{mass}). To estimate
this error we consider the difference between the second and first order
results which gives $\varepsilon _{\mathrm{ASY}}=0.04\,\,\mathrm{GeV}$.
Adding the errors quadratically, we find for the b-quark mass 
\begin{equation}
m_{b}(m_{b})=(4.19\pm 0.05)\,\,\mathrm{GeV}.  \label{result}
\end{equation}

Before going on, a comment on the non perturbative contributions to the two
point correlation function is in order. Using accepted values for the
condensates, we find that their contribution to the b-quark mass for the
considered range of $s_{R}$ is always of the order of $0.001\,\,\mathrm{GeV}$
and, therefore, negligible in comparison to the errors obtained before. We
have also checked that the convergence of the result (\ref{result}) with
respect to the power series expansion in $m^{2}/s$ has no influence to the
error bars in the $s_{R}$ stability domain.

To have a flavor of the whole procedure, we include two figures which
summarize the main steps followed in the calculation of the bottom quark
mass.

\begin{figure}[ht]
\centerline{\psfig{figure=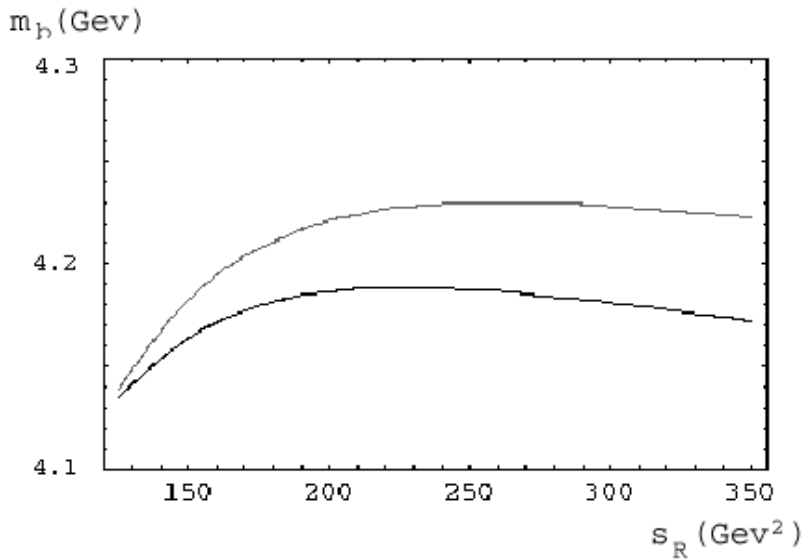,scale=0.7}}
\caption{Stability of the two (upper curve) and three (lower curve) loop
calculations of the bottom quark mass.}
\label{bottomtr}
\end{figure}

As a sample calculation, we have plotted in figure \ref{bottomtr} the
results of $m_{b}^{(1,2)}(m_{b})$ as function of the radius $s_{R}$. The
upper (lower) curve corresponds to the first (second) order in the strong
coupling constant. We find the most stable result for $s_{R}=240\,\,\mathrm{%
GeV}^{2}$ . The difference between second and first order results is $%
\varepsilon _{\mathrm{ASY}}=0.04\,\,\mathrm{GeV}$. The $s_{R}$ dependence is
so tiny that all the values for the mass in the whole range $\left[ 150,350%
\right] \,\,\mathrm{GeV}^{2}$ lie within the relatively small asymptotic
error bar.

\begin{figure}[ht]
\centerline{\psfig{figure=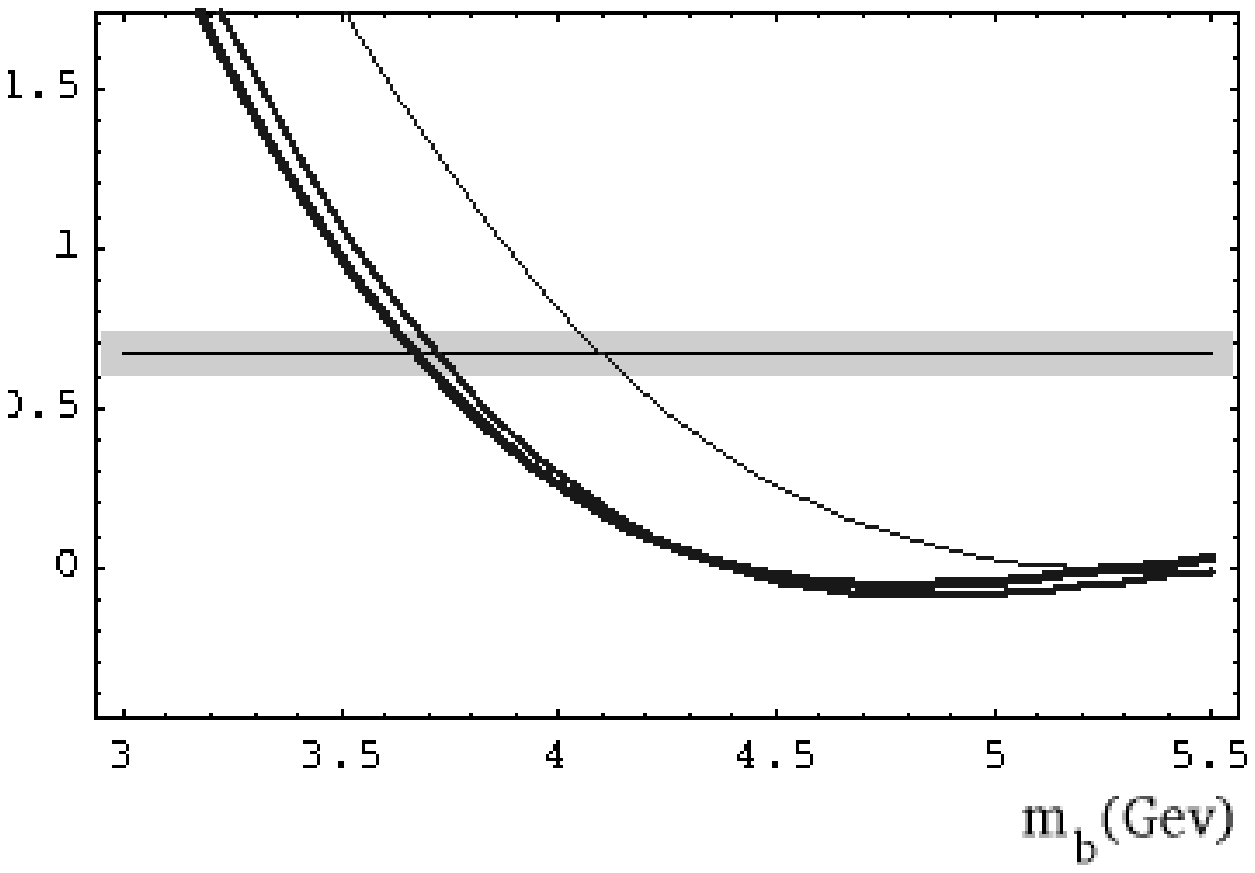,scale=0.8}}
\caption{Results for one (thin), two (medium) and three (thick) loops running
bottom masses at the renormalization and stability scale
$\protect\mu^2=s_{R}=240 \, \, \mathrm{%
GeV}^{2}$.}
\label{bottommass}
\end{figure}

In figure \ref{bottommass} we have plotted the right hand side of the mass
equation (\ref{KEYEQ}) as a function of $m_{b}(\mu )$ with the
renormalization point taken at the value of $s_{R}$ and the latter in the
stability point ($\mu ^{2}=s_{R}=240\,\,\mathrm{GeV}^{2}$). The thin curve
includes only the calculation to one loop, whereas the medium and thick
curves include calculations to two and three loops, respectively. The
straight strip corresponds to the contribution of the experimental data to
the left hand side of equation (\ref{KEYEQ}) including experimental error
bars. Hence, the crossing of the straight strip with the three curves gives
the solution of equation (\ref{KEYEQ}), i.e. the results for the running
mass $m_{b}^{(0,1,2)}(240\,\,\mathrm{GeV^{2}})$. Running this results to $\mu
=m_{b}$ we find the values
quoted in equation (\ref{mass}). Notice the very good convergence of the
perturbative series at the energies relevant to this process.

\section{Conclusions}

The bottom quarks mass has been determined in an unconventional fashion by
employing finite energy QCD sum rules with positive moments. By \ applying
Cauchy's theorem to the $b\overline{b}$ vector current correlator weighted
with a real polynomial, it was possible to virtually eliminate the
contribution of the yet unknown experimental data in the continuum region
from just above the resonances to the beginning of the asymptotic region where
QCD is valid. The experimental input needed is only the resonance masses and
couplings.

The method we use is quite independent of the more popular one based on
inverse moment sum rules. The result we obtain, $m_{b}(m_{b})=(4.19\pm
0.05)\,\,\mathrm{GeV}$, agrees with most of the recent calculations \cite%
{HOANG}, especially with references \cite{PICH} and \cite{KUHN} that study
high and low inverse moment sum rules, respectively. The errors of our
calculation are comparable to the ones quoted in these references. Such
independent determinations of the same quantity serves as a probe of the
validity of the concept of duality which are at the basis of most
calculations in the b-sector.
We find nice convergence of our result with respect to QCD perturbation
theory and negligible non-perturbative contributions.

\end{document}